\documentclass[prl,twocolumn,showpacs,amsmath,amssymb]{revtex4}

\usepackage{graphicx}
\usepackage{dcolumn}
\usepackage{bm}

\begin{document}


\title{Hysteresis behavior in current-driven stationary resonance\\
induced by nonlinearity in the coupled sine-Gordon equation}

\author{Yoshihiko Nonomura}
 \email{nonomura.yoshihiko@nims.go.jp}
\affiliation{%
Computational Materials Science Unit, National Institute for Materials Science, 
Tsukuba, Ibaraki 305-0047, Japan
}%

\date{\today}

\begin{abstract}
Recently novel current-driven resonant states characterized by 
the $\pi$-phase kinks were proposed in the coupled sine-Gordon 
equation. In these states hysteresis behavior is observed with 
respect to the application process of current, and such behavior 
is due to nonlinearity in the sine term. Varying strength of the sine 
term, there exists a critical strength for the hysteresis behavior and 
the amplitude of the sine term coincides with the applied current 
at the critical strength. 
\end{abstract}
\maketitle

{\it Introduction.}
Recently the coupled sine-Gordon equation has been 
intensively studied numerically and analytically as a 
model of THz electromagnetic wave emission from 
intrinsic Josephson junctions (IJJs).~\cite{Ozyuzer,Kadowaki} 
In current-driven emission without external magnetic field, novel 
resonant states characterized by the $\pi$-phase kinks were 
proposed for fairly large surface impedance $Z$,~\cite{Lin2,Koshelev} 
and the present author showed \cite{Nonomura09} that such states 
were stationary for any $Z$ with large enough current.

In these resonant states hysteresis behavior is observed. 
Strong emission which clearly breaks the Ohm's law only takes 
place in the current-increasing process, and the law almost holds 
in the current-decreasing process. Such dynamical behavior is 
due to nonlinearity of the system, namely the sine term in the 
present equation. The main question of the present Letter is 
how such nonlinearity affects on hysteresis behavior in the 
current-driven resonance. May infinitesimal nonlinearity 
cause such behavior, or may there exist a critical value of it?

In order to resolve this question, we introduce an artificial 
parameter $\gamma$ in the equation and verify the strength 
of nonlinearlity continuously. As the parameter $\gamma$ 
decreases from the value in the original equation ($\gamma=1$), 
width of hysteresis also decreases, and there seems to 
exist a nonvanishing critical value $\gamma_{\rm c}$ 
with vanishing hysteresis.
\medskip
\par
{\it Model and formulation.}
When the capacitive coupling is not taken into account, IJJs are 
described by the coupled sine-Gordon equation,~\cite{Koyama}
\begin{equation}
\label{SGeq}
\partial_{x'}^{2}\psi_{l}=(1-\zeta\Delta^{(2)})
\left( \partial^{2}_{t'}\psi_{l}+\beta \partial_{t'}\psi_{l}+\sin \psi_{l}-J' \right),
\end{equation}
with the layer index $l$ and the operator $\Delta^{(2)}$ 
defined in $\Delta^{(2)}X_{l} \equiv X_{l+1}-2X_{l}+X_{l-1}$. 
Quantities are scaled as 
\begin{equation}
x'=x/\lambda_{c},\ t'=\omega_{\rm p}t,\ J'=J/J_{\rm c};\ 
\omega_{\rm p}=c/\left(\sqrt{\epsilon_{\rm c}}\lambda_{c}\right),
\end{equation}
with the penetration depth along the $c$ axis $\lambda_{c}$, the 
plasma frequency in each layer $\omega_{\rm p}$, and the critical 
current $J_{\rm c}$. 
Using material parameters of Bi$_{2}$Sr$_{2}$CaCu$_{2}$O$_{8}$ 
given in Ref.\ \cite{Tachiki}, we result in a large inductive coupling 
$\zeta=4.4\times 10^{5}$, and $\epsilon_{c}=10$ and $\beta=0.02$ are taken.

Neglecting temperature fluctuations and assuming homogeneity 
along the $y$ axis, we have the two-dimensional formula 
(\ref{SGeq}). Following our previous study, \cite{Nonomura09}
width of the junction is chosen as $L_{x}=86\mu$m, and the periodic 
boundary condition (PBC) along the $c$ axis is considered.  Then, 
plasma velocity of the stationary state automatically coincides with 
that of light in IJJs, which corresponds to the case with infinite 
number of junctions, though actual number of junctions $N$ still 
affects on physical properties even in the PBC, especially on 
response to the in-plane magnetic field.~\cite{Nonomura12}
Here we concentrate on the simplest case $N=4$, which takes 
spatial inhomogeneity of the superconducting phase into account.

Since direct evaluation of electromagnetic wave emission 
from edges of a thin sample to vacuum is quite complicated, 
we use a simplified version \cite{Lin08b} of the dynamical boundary 
condition,~\cite{Bulaevskii} where effects outside of the sample are 
only included in the relation between dynamical parts of the rescaled 
electric and magnetic fields, $\tilde{E}'_{l}=\mp Z \tilde{B}'_{l}$, with 
rescaled quantities $E'_{l}$ and $B'_{l}$ related with $\psi_{l}$ as 
$\partial_{t'}\psi_{l}=E'_{l}$ and 
$\partial_{x'} \psi_{l}=(1-\zeta\Delta^{(2)})B'_{l}$, respectively. 
Here we take $Z=30$, which gives strong enough emission 
close to the optimal value.~\cite{Nonomura09} 
The sample along the $x$ axis is divided into $80$ numerical 
grids, and calculations are based on the RADAU5 ODE 
solver.~\cite{radau5} In order to investigate effect of nonlinearity 
in the sine term, Eq.\ (\ref{SGeq}) is slightly modified as 
\begin{equation}
\label{modSG}
\partial_{x'}^{2}\psi_{l}=(1-\zeta\Delta^{(2)})
\left(\partial^{2}_{t'}\psi_{l}+\beta \partial_{t'}\psi_{l}+\gamma \sin \psi_{l}-J'\right),
\end{equation}
and the parameter $\gamma$ is controlled hereafter.
\medskip
\par
\begin{figure}
\includegraphics[height=12.0cm]{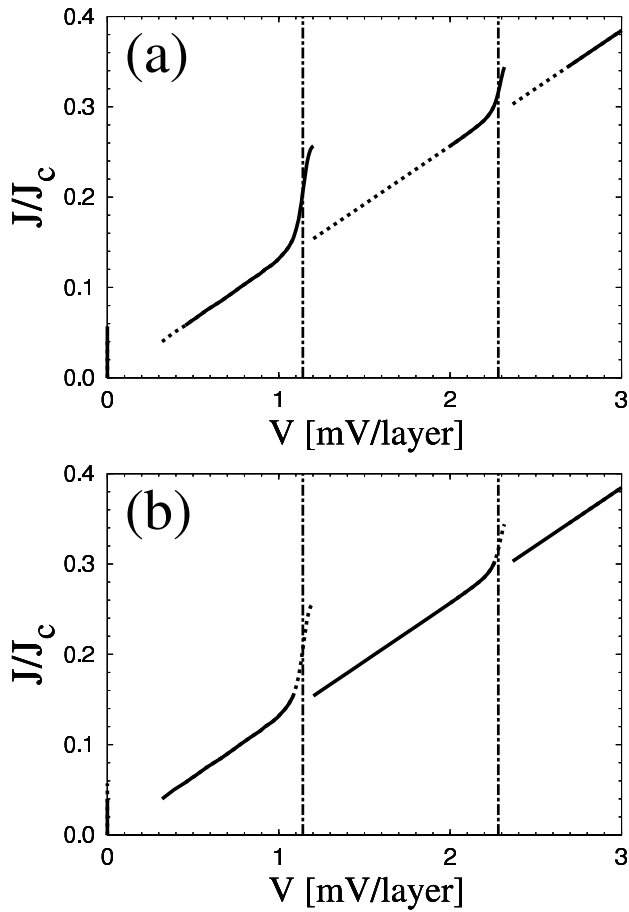}
\caption{\label{fig1}$I$-$V$ curve (solid line) for the 
(a) current-increasing and (b) current-decreasing processes. 
Dotted line stands for the curve in another figure, and the 
solid-dotted lines denote the voltages corresponding to 
the cavity resonance points (\ref{cavV}).}
\end{figure}

{\it Hysteresis in the original equation.}
First, the $I$-$V$ curve of the original equation (\ref{SGeq}) is displayed 
in Figs.~\ref{fig1}(a) and \ref{fig1}(b). As long as the fundamental and 
first harmonic modes are observed, large hysteresis is quite apparent. 
In the current-increasing process (Fig.\ \ref{fig1}(a)), 
the current rapidly increases as the voltage approaches 
the value corresponding to the cavity resonance point 
given by the ac Josephson relation (broken lines), 
\begin{equation}
  \label{cavV}
  V=\phi_{0}f=\phi_{0}\frac{c}{\sqrt{\epsilon_{c}}}\frac{n}{2L_{x}}
  \approx 1.14\ n\ [{\rm mV}],
\end{equation}
with the number of nodes $n$($=1$: the fundamental mode). 
In the vicinity of emission peaks, the voltage exceeds the value 
given by Eq.~(\ref{cavV}), which assumes perfect cavity resonance. 
In experiments about 10\% of frequency looks tunable in a single 
resonant branch by varying the current,~\cite{Tsujimoto,Benseman} 
which is consistent with this result. When a much larger value of 
$Z$ is taken, the range of voltage becomes much smaller \cite{Lin2} 
during varying the current in the region of strong emission. 
\begin{figure}
\includegraphics[height=12.0cm]{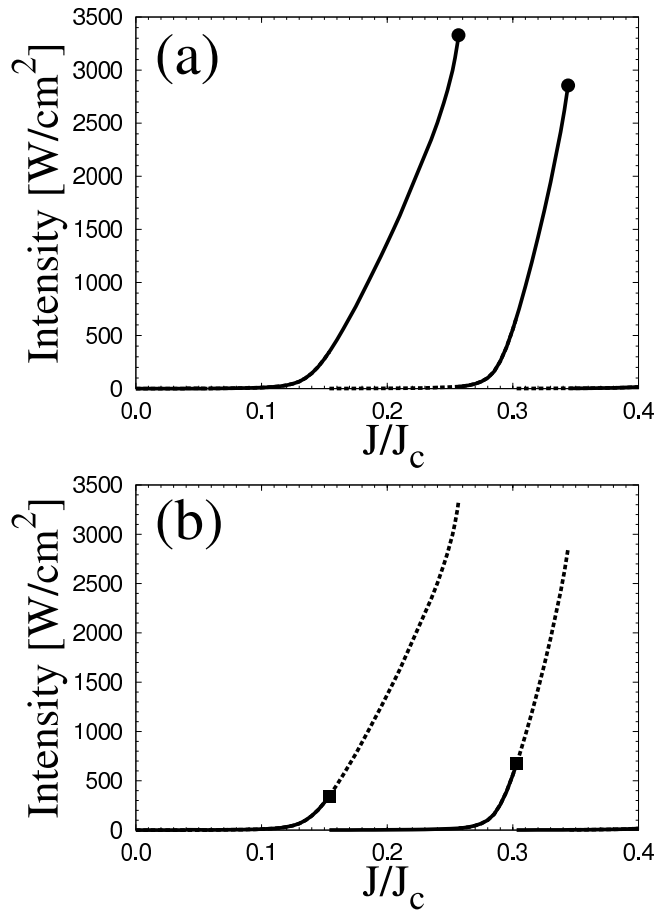}
\caption{\label{fig2}Emission intensity versus current for the 
(a) current-increasing and (b) current-decreasing processes. 
Dotted line stands for the curve in another figure. Circles and 
squares represent the emission peak in each node and in the 
current-increasing and current-decreasing processes, respectively.}
\end{figure}

Emission intensity corresponding to the situations 
for Figs.~\ref{fig1}(a) and \ref{fig1}(b) is plotted versus 
current in Figs.\ \ref{fig2}(a) and \ref{fig2}(b), respectively. 
Strong emission observed in the current-increasing process 
is quite reduced in the current-decreasing process. 
Especially in the fundamental mode, the maximum intensity 
in the current-decreasing process is almost one order 
smaller than that in the current-increasing process. In such a 
case emission is expected to be invisible when experimental 
noise is overloaded in the reverse (current-decreasing) process, 
which may represent ``irreversible" emission in experiments.
\medskip
\par
\begin{figure}
\includegraphics[height=12.0cm]{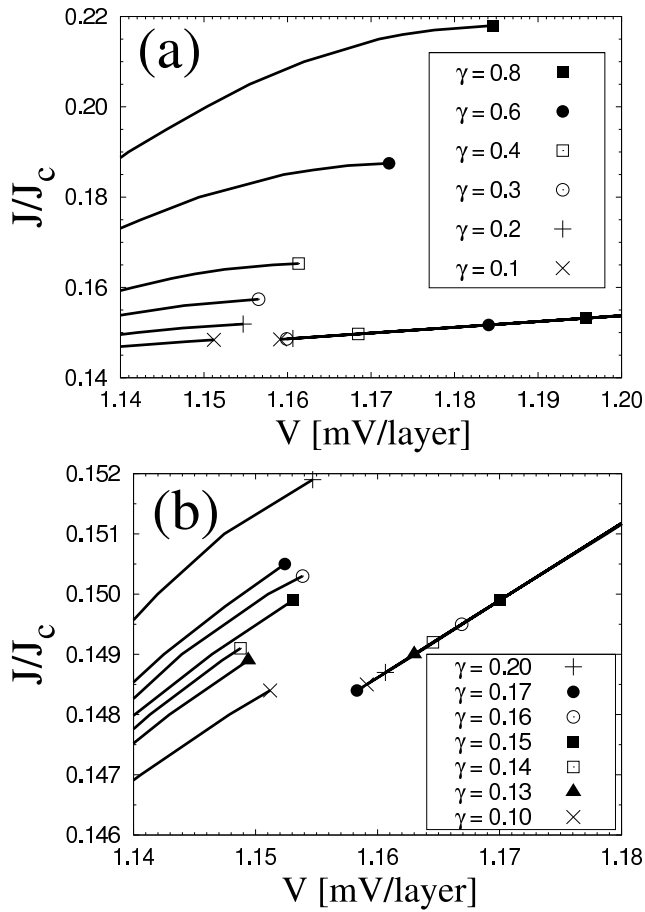}
\caption{\label{fig3}$I$-$V$ curve for various values of $\gamma$ 
for (a) $\gamma=0.8$ to $0.1$ and (b) $\gamma=0.2$ to $0.1$ 
(expanded figure around the irreversible-reversible boundary). 
Edges of the $n=1$ curve in the current-increasing process and 
the $n=2$ curve in the current-decreasing process are visualized 
by various symbols.}
\end{figure}

{\it Hysteresis in the equations with varying nonlinearity.}
Next, the $I$-$V$ curve of the modified equation (\ref{modSG}) is 
investigated. Here we concentrate on hysteresis behavior around 
the emission peaks in the fundamental ($n=1$) mode, and only 
observe the $n=1$ curves in the current-increasing process near 
the upper edges and the $n=2$ curves in the current-decreasing 
process near the lower edges. In Fig.~\ref{fig3}(a), parameter 
dependence of the upper and lower edges is visualized by 
various symbols for a wide range of parameters between 
$\gamma=0.8$ and $0.1$. As $\gamma$ decreases, 
the width of hysteresis, namely the difference of the 
voltages at the both edges in the current-varying processes, 
becomes smaller and smaller. Hysteresis is observed up to 
$\gamma=0.2$, while it is invisible at $\gamma=0.1$.

Then, more precise measurement is made between $\gamma=0.2$ 
and $0.1$ as shown in Fig.~\ref{fig3}(b). Although the upper edges 
of the $n=1$ curve decrease monotonically, the lower edges of the 
$n=2$ curve exhibit non-monotonic behavior. The lower edges 
decrease up to $\gamma=0.17$, increase up to $\gamma=0.15$, 
and decrease again for smaller $\gamma$. Hysteresis behavior 
is observed up to $\gamma=0.16$, and from $\gamma=0.15$ 
current at the upper and lower edges is the same, 
or the current-varying process becomes reversible.
\medskip
\par
{\it Discussions.}
These results suggest that the boundary between the irreversible 
and reversible behavior locates between $\gamma=0.16$ and 
$0.15$, but further precise calculation in search for the next digit 
of $\gamma$ may not be productive. Instead, we point out that 
the current at $\gamma=0.15$ is $J'=0.150$ on the edges, or 
that the amplitude of the sine term coincides with the constant 
term at $\gamma=0.15$ in the equation (\ref{modSG}). 
This fact strongly suggests that $\gamma=0.15$ is the 
irreversible-reversible boundary. Quite recently reversible 
THz wave emission from IJJs was reported,~\cite{Minami} 
which may be related with the present finding.

Introduction of the parameter $\gamma$ can be regarded as modification 
of the critical current, namely  $J_{\rm c} \rightarrow \gamma J_{\rm c}$. 
Then, $\gamma=J'$ means that the current $J$ is equal to the modified 
critical current $\gamma J_{\rm c}$, where superconductivity breaks 
down and the present model is not justified anymore. It is natural 
that hysteresis behavior of the model may also change there. 
In the original equation (\ref{SGeq}), the situation $J>J_{\rm c}$ 
occurs in higher harmonic modes and similar behavior may 
also be observed. Numerical study along this direction is 
now in progress.~\cite{Nonomura12b} 

Frequency of resonance $f$ in Eq.\ (\ref{cavV}) is derived from the 
linearized version of Eq.\ (\ref{SGeq}),~\cite{Sakai} and the role 
of nonlinearity is to generate the $\pi$-phase kinks \cite{Lin2} and 
hysteresis behavior. The present result shows that the hysteresis 
behavior is also nontrivial. We consider this behavior is general 
and can be regarded as a prototype of hysteresis in 
harmonic oscillations induced by nonlinearity.
\medskip
\par
{\it Summary.}
In the present Letter we numerically investigate characteristic 
behavior of novel current-driven resonant states in the coupled 
sine-Gordon equation. These states were proposed theoretically 
as strong emission states in THz electromagnetic wave emission 
from intrinsic Josephson junctions, and such emission depends 
on the application process of current. That is, hysteresis is 
observed in emission behavior driven by applied current. 
Such irreversible behavior is due to nonlinearity of the 
equation, and we vary the strength of the sine term.

As long as the fundamental 
resonant mode is observed, the maximum intensity of emission 
(and consequently the maximum value of the current) decreases 
monotonically as the strength of nonlinearity decreases, and at 
the nonvanishing critical value hysteresis behavior disappears 
and the amplitude of the sine term coincides with that of the 
applied rescaled current at the emission peak. From a physical 
point of view, superconductivity and hysteresis behavior vanish 
at the same time.
\medskip
\par
{\it Acknowledgments.}
The present work was partially supported by Grant-in-Aids for 
Scientific Research (C) No.\ 20510121 from JSPS.

\end{document}